\documentclass[prl,twocolumn,superscriptaddress]{revtex4}
\usepackage[dvips]{epsfig}
\usepackage{graphicx,amsmath,amssymb}
\usepackage{amsmath,amssymb,lscape,float}
\usepackage{hyperref}

    \def\<{\langle}         \def\>{\rangle}

  \def\V0{{\mathbf 0}}

  \def\B0{{\mathbf 0}}

\def\be{\begin{equation}}       \def\ee{\end{equation}}
\def\bea{\begin{eqnarray}}      \def\eea{\end{eqnarray}}

\begin{document}

\title{Incommensurate superfluidity of bosons 
       in a double-well optical lattice }

\author{Vladimir M. Stojanovi\'c}
\affiliation{Department of Physics, Carnegie Mellon University,
Pittsburgh, Pennsylvania 15213, USA}
\author{Congjun Wu}
\affiliation{Department of Physics, University of California, 
San Diego, California 92093, USA}
\author{W. Vincent Liu
}
\affiliation{Department of Physics and Astronomy, University of
Pittsburgh, Pittsburgh, Pennsylvania 15260, USA}
\author{S. Das Sarma}
\affiliation{Condensed Matter Theory Center, Department of Physics, 
             University of Maryland, College Park, Maryland 20742, USA}

\date{\today}
\begin{abstract}
{We study bosons in the first excited Bloch band of a double-well
optical lattice, recently realized at NIST.  By calculating the relevant
parameters from a realistic nonseparable lattice
potential, we find that in the most favorable cases the boson lifetime
in the first excited band can be several orders of magnitude longer
than the typical nearest-neighbor tunnelling timescales, in
contrast to that of a simple single-well lattice. In addition, for
sufficiently small lattice depths the excited band has minima at
nonzero momenta incommensurate with the lattice period, which opens a
possibility to realize an exotic superfluid state that spontaneously breaks
the time-reversal, rotational, and translational symmetries. We
discuss possible experimental signatures of this novel state.}
\end{abstract}

\pacs{03.75.Lm, 05.30.Jp, 67.85.Hj} 
\maketitle

Optical lattices provide an exquisite tool for controlled exploration
of novel types of order in cold atomic gases~\cite{optlattreviews}. 
In particular, realization of a (quasi) two-dimensional (2D) 
double-well (DW) optical lattice at NIST was one of the latest 
major developments in the experimental cold-atom physics~\cite{NISTdbwell},
motivating further experimental~\cite{FoelingNature:07} and 
theoretical~\cite{Danshita} efforts. 

Ultracold atoms, either fermionic or bosonic, in 
higher Bloch bands have recently ignited a great deal of 
interest~\cite{Isacsson:05,Scarola:05,Alon:05,Kohl+:05,Liu+Wu:06,
Ho:06,Mueller+:07,Congjuns,Zhao+Wu:08}.
In the fermion case Pauli blocking enables one to populate high 
Bloch bands by simply increasing the atomic density~\cite{Kohl+:05}.
By contrast, the true ground-state condensation of bosons
only occurs in the lowest $s$-orbital band even for high boson densities.
When the majority of bosons populates a higher band, bosons
are in excited states with a finite lifetime.
Isacsson {\it et al.} \cite{Isacsson:05} have studied bosons in
the first excited band of 1D optical lattice and found lifetimes that 
are on the order of $10-100$ times longer than the typical nearest-neighbor 
tunnelling time, a prediction that has recently been experimentally 
corroborated by the Mainz group~\cite{Mueller+:07}. 
In this Letter, we study weakly-interacting
cold bosons populating the first excited band of a quasi-2D
DW optical lattice.

We show that in the superfluid regime Bose-Einstein condensation (BEC) takes 
place at an incommensurate nonzero momentum, which spontaneously breaks 
time-reversal, rotational, and translational symmetries. We further demonstrate
that, due to vastly reduced available phase space for the decay to 
the lowest band, the lifetime of a Bose gas in the first excited band of a 
DW lattice can be several orders longer
than the inverse tunnelling rate, which in turn sets the
characteristic time needed to establish phase coherence 
in the system~\cite{Clark+Jaksch:04}.

The DW lattice consists of a nonseparable optical
potential in the $x$-$y$ plane and a conventional optical 
potential in the $z$-direction. In the case with DWs oriented in the 
$x$-direction, in a coordinate system with the origin at a maximum 
point of the ``in-plane''-lattice light
intensity, the optical potential in the $x$-$y$ plane is given 
by~\cite{NISTdbwell}
\begin{eqnarray}
\label{potent}
V(x,y)&=& 2V_{0}\big\{[\cos(2k_{\textrm{\tiny{L}}}y)
-\cos(2k_{\textrm{\tiny{L}}}x) ] \nonumber \\
&+& 2r [\cos(k_{\textrm{\tiny{L}}}x)
+\cos(k_{\textrm{\tiny{L}}}y) ]^{2}\big\} \:.
\end{eqnarray}
\noindent 
Here $k_{\textrm{\tiny{L}}}=2\pi/\lambda$ is the magnitude of the 
laser wave-vector, $V_{0}=-|V_{0}|<0$ (red-detuned lattice), 
while $r\equiv I_{z}/I_{xy}$ stands for the relative intensity of two
components of light with the in-plane and out-of-plane 
polarizations.
The band structure of the 
single-particle Hamiltonian $H_{0}=-(\hbar^{2}/2m_{b})(\partial_{x}^{2}
+\partial_{y}^{2})+V(x,y)$ ($m_{b}$--boson mass) in the $x$-$y$ plane 
can be solved by using the plane-wave basis.
The corresponding matrix elements of $H_0$ read
\begin{eqnarray}
&&\langle\mathbf{k}+\mathbf{G}_{m,n}\:|H_{0}|\:\mathbf{k}
+\mathbf{G}_{m,n}\rangle = 
E_{\scriptscriptstyle R}\big\{[(k_{x}/k_{\textrm{\tiny{L}}})
+m+n]^{2} \nonumber\\
&&+[(k_{y}/k_{\textrm{\tiny{L}}})+m-n]^{2}\big\} \:, 
\nonumber \\
&&\langle \mathbf{k}\:|H_{0}|\:\mathbf{k}
+\mathbf{G}_{\pm 1,0}\rangle =\langle \mathbf{k}\:|H_{0}|\:\mathbf{k}
+\mathbf{G}_{0,\pm 1}\rangle = 2rV_{0} \nonumber \:,\\
&&\langle \mathbf{k}\:|H_{0}|\:\mathbf{k}
+\mathbf{G}_{\pm 1,\pm 1}\rangle= (r-1)V_{0} \:,\\
&&\langle \mathbf{k}\:|H_{0}|\:\mathbf{k}
+\mathbf{G}_{\pm 1,\mp 1}\rangle = (r+1)V_{0} \nonumber \:,
\end{eqnarray}  
\noindent where $\mathbf{k}=(k_{x},k_{y})$ is the wave-vector,
$\mathbf{G}_{m,n}\equiv m\mathbf{b}_{1}+n\mathbf{b}_{2}$
($m,n$--integers) are the reciprocal-lattice vectors (with basis  
$\mathbf{b}_{1,2}=k_{\textrm{\tiny{L}}}(\hat{\mathbf{e}}_{x}
\pm\hat{\mathbf{e}}_{y})$), and $E_{\scriptscriptstyle R}=\hbar^{2}
k_{\textrm{\tiny{L}}}^{2}/(2m_{b})$
is the recoil energy. 

\begin{figure}[htb]
\begin{center}
\includegraphics[width=0.7\linewidth]{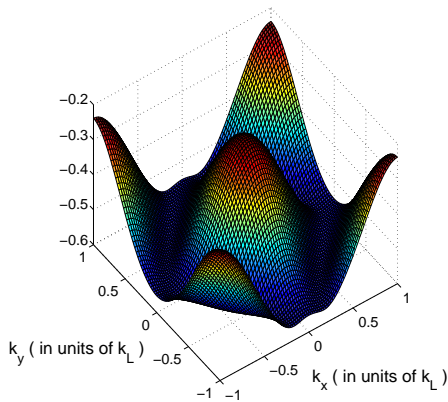}
\end{center}
\caption{Excited-band dispersion for 
$V_{0}/E_{\scriptscriptstyle R}=-1.0$, $r=0.08$.} \label{excsurf}
\end{figure}

\begin{figure}[t!]
\begin{center}
\includegraphics[width=0.55\linewidth]{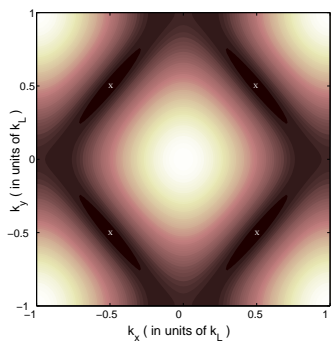}
\end{center}
\caption{Contour plot of the excited-band dispersion for 
$V_{0}/E_{\scriptscriptstyle R}=-1.0$, $r=0.08$. Band-minima are marked 
by 'x'.} \label{exccontour}
\end{figure}

The dispersion of the first 
excited band is depicted in Fig.~\ref{excsurf}, where 
the energies are expressed in units of $E_{\scriptscriptstyle R}$.
\noindent 
While the lowest band has minimum at $\mathbf{k}=0$, the first
excited band has maximum at $\mathbf{k}=0$ and minima for $\mathbf{k}\neq 0$.
For larger values of $|V_{0}|$, these minima occur at commensurate
wave-vectors of
$\mathbf{k}=(\pm 1,0)k_{\textrm{\tiny{L}}}$ and 
$(0,\pm 1)k_{\textrm{\tiny{L}}}$.
However, for optical lattice depths smaller (in absolute value) than some $r$-dependent
threshold value, i.e., for $|V_{0}|<V_{\textrm{th}}(r)$, these minima occur at
wave-vectors $\mathbf{K}$ that are incommensurate with the lattice period in both 
$x$- and $y$-directions and independent of $V_{0}$: 
$K_{\textrm{x}}=\pm 0.5\:k_{\textrm{\tiny{L}}}$, 
$K_{\textrm{y}}=\pm 0.5\:k_{\textrm{\tiny{L}}}$.
For example, the threshold value for $r=0.08$ is $V_{\textrm{th}}
\approx 1.08\:E_{\scriptscriptstyle R}$, while for $r=0.15$ it is $V_{\textrm{th}}
\approx 0.83\:E_{\scriptscriptstyle R}$. Contour plot of the excited-band dispersion for 
$V_{0}/E_{\scriptscriptstyle R}=-1.0$ and $r=0.08$ is displayed in Fig. \ref{exccontour}.
The band-minima are only two-fold degenerate, related by a mirror symmetry,
because $(\pm\frac{1}{2} k_{\textrm{\tiny{L}}},\pm\frac{1}{2} k_{\textrm{\tiny{L}}})$ 
are equivalent to each other and so are 
$(\pm\frac{1}{2} k_{\textrm{\tiny{L}}}, \mp\frac{1}{2} k_{\textrm{\tiny{L}}})$.

The two relevant bands originate from the lowest-energy 
states of a DW: the lowest (+) band from the DW ground
state (even parity), the excited (--) band from the first excited
state (odd parity) of a DW. These two states can be sought in the form of
``bonding'' and ``anti-bonding'' linear combinations of single-well
Gaussians, respectively, similar to the
Heitler-London variational approach to the $H_{2}^{+}$ molecular 
ion. For the DW comprising potential-minima at $(x_{1},0)$ 
and $(x_{2},0)$, 
\begin{equation}
\Phi_{\pm}(x,y) =\frac{\varphi(x-x_{1},y)
\pm\varphi(x-x_{2},y)}{\sqrt{2(1\pm S)}} \:, \label{twofunc1}
\end{equation}
\noindent where $\varphi(x,y)=(\pi \sigma^{2})^{\scriptscriptstyle -1/2}
\:e^{-\frac{x^{2}+y^{2}}{2\sigma^{2}}}$ is a 2D Gaussian and 
$S=\int\varphi^{*}(x-x_{1},y)\varphi(x-x_{2},y)
\:d^{2}\mathbf{r}=e^{-b^{2}/(4\sigma^{2})}$ is the overlap integral 
of two such Gaussians, a distance $b\equiv x_{2}-x_{1}$ apart. 
Optimal value $\sigma_{\scriptscriptstyle 0}$ of the Gaussian-width 
$\sigma$ is obtained by minimizing the expectation value 
$\langle\Phi_{+}(x,y)|H_{0}|\Phi_{+}(x,y)\rangle$ 
of the DW ground-state energy over this parameter. 
The value of $\sigma_{\scriptscriptstyle 0}$ becomes
larger with decreasing lattice depth (i.e., for decreasing $|V_{0}|$).
In particular, our calculation shows that for 
$|V_{0}|/E_{\scriptscriptstyle R}\lesssim 1.0$ 
one has $\sigma_{\scriptscriptstyle 0}\gtrsim 0.39\:b$.  
Strictly speaking, the Wannier functions are
well described by the ansatz in Eq. \eqref{twofunc1} only for 
not-too-small $|V_{0}|$ (implying that the mid-barrier between single
wells is sufficiently high). By comparing the variational ground state
energies of a DW corresponding to different lattice depths 
with band-structure calculations, we
can estimate that this approach is accurate for 
$|V_0|\gtrsim 0.4\:E_{\scriptscriptstyle R}$,
which puts the lower bound on the lattice depths 
that will hereafter be discussed.

The $z$-dependent part of the full 3D Wannier functions 
$\Phi_{\pm}(x,y,z)=\Phi_{\pm}(x,y)\phi(z)$ takes on the 
standard Gaussian form
$\phi(z)=(\pi\xi_{z}^{2})^{\scriptscriptstyle -1/4} \:
e^{-\frac{z^{2}}{2\xi_{z}^{2}}}$, 
where $\xi_{z}$ is the effective harmonic ``zero-point'' length in the
$z$-direction, characterizing the transverse confinement of the system.

\begin{figure}[htb]
\begin{center}
\includegraphics[width=0.6\linewidth]{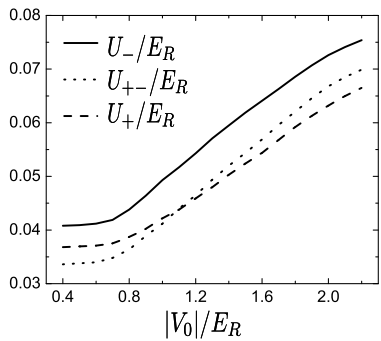}
\end{center}
\caption{Hubbard interaction parameters (in units of $E_{\scriptscriptstyle R}$) 
for $r=0.08$, $a_{s}/\xi_{z}=0.06$, and different lattice depths.} 
\label{HubbardU}
\end{figure}

The intra-band and inter-band Hubbard interaction parameters are
given by $U_{\pm}=(g/\xi_{z}\sqrt{2\pi})\int\Phi_{\pm}^{4}(x,y)
d^{2}\mathbf{r}$ and $U_{+-} = (g/\xi_{z}\sqrt{2\pi})\int
\Phi_{+}^{2}(x,y)\Phi_{-}^{2}(x,y)d^{2}\mathbf{r}$, respectively,
where $g \equiv 4\pi\hbar^{2}a_{s}/m_{b}$. Calculation yields
\begin{eqnarray}
\frac{U_{+-}}{E_{\scriptscriptstyle R}} &=&
\frac{2(a_{s}/\xi_{z})}{\sqrt{2\pi}(\sigma_{\scriptscriptstyle 0} 
k_{\textrm{\tiny{L}}})^{2}} \:,\label{upm} \\
\frac{U_{\pm}}{U_{+-}} &=&
\frac{1+3e^{-b^{2}/2\sigma_{\scriptscriptstyle 0}^{2}}
\pm 4e^{-3b^{2}/8\sigma_{\scriptscriptstyle 0}^{2}}}
{(1\pm e^{-b^{2}/4\sigma_{\scriptscriptstyle 0}^{2}})^{2}} \: .
\end{eqnarray}
\noindent These interaction energies are proportional to 
the ratio $a_{s}/\xi_{z}$, the realistic values of which can be estimated 
to be between $0.03$ and $0.07$~\cite{NISTdbwell}. Evaluation of 
Hubbard $U$'s, displayed in Fig.~\ref{HubbardU}, shows
that they are smaller than the bandgap between the lowest and the
excited band for $|V_{0}|\lesssim 2.2\: E_{\scriptscriptstyle R}$, the 
latter being the largest optical lattice depth that we shall be concerned with
in what follows. [It is useful to note, however, that parameter $V_{0}$
here does not have entirely the same meaning as
lattice depth in the case of conventional optical lattices.] 
For instance, for $V_{0}/E_{\scriptscriptstyle R}=-1.0$, $r=0.08$, and 
$a_{s}/\xi_{z}=0.06$, values of Hubbard parameters are 
$U_{+}=0.0426\:E_{\scriptscriptstyle R},
\:U_{-}=0.0498\:E_{\scriptscriptstyle R},
\:U_{+-}=0.0420\:E_{\scriptscriptstyle R}$.
Unlike the situation in ordinary (single-well) optical 
lattices, where the Hubbard energy of the excited 
($p$) band is smaller than that of the lowest ($s$) band,
here we find that $U_{-}>U_{+}$. 

While the `$-$'-band minima are degenerate, condensate 
fragmentation~\cite{E.Mueller:06} in momentum space is prevented 
by the presence of interactions, no matter how weak.
It is also easy to check that condensation into any state 
that is a linear superposition of two different band minima has higher 
energy than in a single minimum.
Therefore, our state is a ``simple'' BEC~\cite{Leggettbook} 
at wave-vector denoted with $\mathbf{K}$. It is 
interesting to note that `$-$'-band has nearly flat dispersion around 
the minima, which may pose a challenge to the experimental realization 
of this state.

Our primary objective is to study the influence of interactions on 
the lifetime of bosons in a band minimum of the excited band.
Thus it is essential to elucidate the dominant decay process involved.
It is worthwhile noting that the DW-type systems have 
asymmetric level-spacings, with energy levels appearing in pairs and the 
energy gap between the `$-$'-band and the second-excited band ($W_{32}$) 
being much greater than the one between the `$+$'- and `$-$'-bands 
($W_{21}$): for instance, for $V_{0}/E_{\scriptscriptstyle R}=-1.0$ and $r=0.08$ our 
band-structure calculation yields $W_{32}=1.52\:E_{\scriptscriptstyle R}$, 
$W_{21}=0.079\:E_{\scriptscriptstyle R}$. Therefore, 
promotion of particles from the `$-$' band to the yet higher one, 
relevant for the decay channel studied in Ref.~\cite{Isacsson:05}, 
is energetically very costly. Thus the dominant decay process 
we need consider is the one where two bosons decay from 
the `$-$'-band-minimum to the `$+$'-band. The increase in 
the interaction energy, which is predominantly due to 
the exchange-energy contribution (inter-band interactions, here 
described by $U_{+-}$), compensates for the decrease in the 
single-particle band energy. 

We recall the expression for the interaction energy 
\bea
\displaystyle V_{\textrm{int}}=
\frac{1}{2N_0}\sum_{\scriptscriptstyle n,m;\mathbf{k,k',q}}
U_{m^\prime n^\prime, mn}a_{n^\prime,\mathbf{k+q}}^{\dagger}
a_{m^\prime,\mathbf{k'-q}}^{\dagger} 
a_{m,\mathbf{k'}}
a_{n,\mathbf{k}}\:,
\eea
where $m,n,m^\prime,n^\prime$ are band indices, and $N_0$ is the number 
of unit cells in the system.
For the first two bands, interaction parameters are
$U_{++,++}=U_+$, $U_{--,--}=U_-$ and
$U_{++,--}=U_{--,++}=U_{+-,-+}=U_{+-,+-}=U_{-+,-+}=U_{-+,+-}
=U_{+-}$.
Let us assume that the initial state is
$|\psi_i\rangle = (a^\dagger_{-, \bf K})^N |\textrm{vac}\rangle$,
where $N$ is the number of particles in the condensate
and $|\textrm{vac}\rangle$ stands for the boson vacuum,
while the final state is $|\psi_f\rangle= 
a^\dagger_{+,\bf k_1} a^\dagger_{+,\bf k_2} (a^\dagger_{-, \bf K})^{N-2} 
|\textrm{vac}\rangle$.
The change of the interaction energy between $|\psi_i\rangle$ and
$|\psi_f\rangle$ is 
$\Delta E_{\textrm{int}}= E_{\textrm{int},f}-E_{\textrm{int},i}
= 4 \nu U_{+-} - 2 \nu U_{-} \:,$
where $\nu=N/N_0$ is the average filling per DW.
The energy conservation condition reads 
$\Delta E_{\textrm{int}}+\Delta E_{\textrm{kin}}=0$, i.e.,
\bea
2\varepsilon_{-}(\bf{K})-\varepsilon_{+}(\bf {k_1})
-\varepsilon_{+}(\bf{k_2})= 4\nu U_{+-}-2\nu U_{-} \:.
\label{energcons}
\eea

By employing the Fermi Golden Rule and the tight-binding
condition 
in the $z$-direction, in which the system is tightly confined, 
we arrive at the expression for the transition rate per DW:
\begin{equation}\label{damprate}
w=\frac{1}{\hbar}\left(\frac{4\pi\nu \hbar^{2}a_{s}}
{m_{b}\xi_{z}}\right)^{2}\sum^\prime_{\mathbf{k_{1}},\mathbf{k_{2}}}
\frac{\big|\int d^{2}\mathbf{r}\:\Psi_{+,\mathbf{k_{1}}}^{*}
\Psi_{+,\mathbf{k_{2}}}^{*}\Psi_{-,\mathbf{K}}^{2}
\big|^{2}}{\rho(\varepsilon_{+}(\mathbf{k_{1}}))^{-1}
+\rho(\varepsilon_{+}(\mathbf{k_{2}}))^{-1}} \:.
\end{equation}
\noindent Here $\Psi_{\pm,\mathbf{k}}(\mathbf{r})=e^{i\mathbf{k}
\cdot\mathbf{r}}u^{\pm}_{\mathbf{k}}(\mathbf{r})$ are the Bloch
wave-functions for the two bands ($u^{\pm}_{\mathbf{k}}(\mathbf{r})$
being their corresponding lattice-periodic parts), $\rho(\varepsilon)=
\sum_{\mathbf{k}\in\textrm{B.Z.}}\delta(\varepsilon-\varepsilon_{+}(\mathbf{k}))$  
is the density-of-states for the `$+$'-band, 
and the prime indicates that the momentum summation is restricted to the 
pairs ($\mathbf{k_{1}},\mathbf{k_{2}}$) that satisfy both the energy-conservation 
condition in Eq.~\eqref{energcons} and momentum conservation up to a
reciprocal lattice vector ($\mathbf{k_{1}}+\mathbf{k_{2}}=2\mathbf{K}+\mathbf{G}$).
The transition rate is calculated by way of numerical evaluation of
the `$+$'-band 
density-of-states $\rho(\varepsilon)$ and the spatial integral  
$\int d^{2}\mathbf{r}\:\Psi_{+,\mathbf{k_{1}}}^{*}
\Psi_{+,\mathbf{k_{2}}}^{*}\Psi_{-,\mathbf{K}}^{2}$ in Eq.~\eqref{damprate}.
The latter is based on the eigenvectors obtained in the 
band-structure calculation, that is, coefficients in the 
Fourier expansion $u^{\pm}_{\mathbf{k}}(\mathbf{r})=
\sum_{\mathbf{G}}C^{\pm}_{\mathbf{k},\mathbf{G}}
\:e^{i\mathbf{G}\cdot\mathbf{r}}$ over reciprocal-lattice vectors.

\begin{figure}[htb]
\begin{center}
\includegraphics[width=0.75\linewidth]{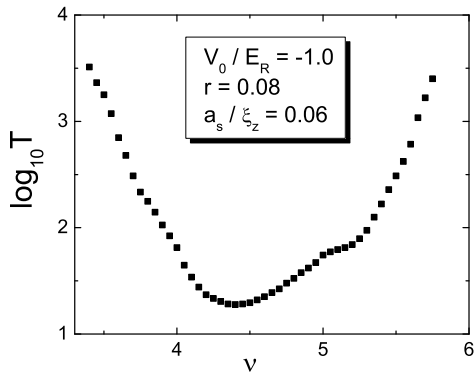}
\end{center}
\caption{Boson lifetime in the first excited band for different
filling factors $\nu$. Values of parameters are indicated.} \label{LifetimeDW}
\end{figure}

To quantify the stability of bosons in the excited band, it is
pertinent to compare it with the timescale corresponding to the
hopping amplitude $t_{h}= \left\langle\Phi_{-}^{i}\left|H_{0}\right
|\Phi_{-}^{i+1}\right\rangle$, where $\Phi_{-}^{i},\:\Phi_{-}^{i+1}$
are the Wannier functions corresponding to a pair of adjacent (in the
$x$-direction) DWs. For $V_{0}/E_{\scriptscriptstyle R}=-1.0$, $r=0.08$ we obtain
$t_{h}=-0.1007\:E_{\scriptscriptstyle R}$. With the choice of $^{87}$Rb atoms 
and $\lambda=810\:$nm ($E_{\scriptscriptstyle R}\approx h\times 3.5\:$kHz), the 
hopping timescale is $\hbar/|t_{h}|\approx 0.45\:$ms. The logarithm of the 
dimensionless lifetime $T=|t_{h}|/(\hbar w)$, depicted 
with varying filling factor in Fig.~\ref{LifetimeDW}, exhibits
non-monotonous dependence on the filling. The salient feature is
that the lifetime is very long for fillings smaller than some critical
value and also for fillings larger than some higher critical
value. This is easy to understand from the energy-conservation
requirement: namely, the change in band-energy in the decay process of
interest is bounded both from above (by
$2(\varepsilon_{-}^{\textrm{min}}-\varepsilon_{+}^{\textrm{max}})$)
and from below (by
$2(\varepsilon_{-}^{\textrm{min}}-\varepsilon_{+}^{\textrm{min}})$),
while the change in the interaction energy is linear in the filling
factor (recall Eq.~\eqref{energcons}).

\begin{figure}[htb]
\begin{center}
\includegraphics[width=0.7\linewidth]{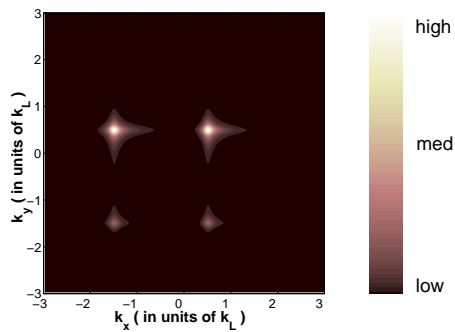}
\end{center}
\caption{Prediction of the density distribution in a time-of-flight
experiment, for $V_{0}/E_{\scriptscriptstyle R}=-1.0$, $r=0.08$, and
condensate wave-vector $\mathbf{K}=(0.5,0.5)k_{\textrm{\tiny{L}}}$.} \label{tofimage}
\end{figure}

The predicted superfluid phase can be experimentally 
identified from the time-of-flight density distribution 
$\langle n(\mathbf{r})\rangle_{t} \propto
\sum_{\mathbf{G}} |\tilde{\Phi}_{-}(\mathbf{k})|^{2}
\:\delta^{\scriptscriptstyle 2}
(\mathbf{k}-\mathbf{K}-\mathbf{G})$, where
$\mathbf{k}=m_{b}\mathbf{r}/(\hbar t)$ and $\tilde{\Phi}_{-}(\mathbf{k})$
is the Fourier transform of the Wannier function $\Phi_{-}(x,y)$. The resulting 
density distribution, displayed in Fig. \ref{tofimage}, has Bragg peaks at the 
condensate wave-vector $\mathbf{K}$ and some of the wave-vectors 
related to it by a translation through a reciprocal-lattice vector. 
Importantly, this distribution is asymmetric with respect to both $\mathbf{k=K}$ 
and $\mathbf{k=0}$. This is a combined effect of the odd parity of 
the Wannier function $\Phi_{-}(x,y)$ and its extended character in real space (that is,
localized in momentum space) in the regime of interest.

The resulting many-body state $|\Psi\rangle$ spontaneously 
breaks the time-reversal and lattice translational symmetries, 
with circulating bond currents~\cite{Paraoanu:03} $\mathcal{J}_{ij}\propto\langle 
\Psi|a^{\dagger}_{i}a_{j}-a^{\dagger}_{j}a_{i}|\Psi\rangle\:\propto
\sin(\mathbf{K}\cdot\mathbf{R}_{ij})$ between sites $i$ and $j$ 
with relative position vector $\mathbf{R}_{ij}$. Unlike the supersolid 
phase~\cite{Scarola:05}, which has the superfluid and density-wave orders,
our state besides the superfluid order supports a current order. This state
does not carry a net current, since the group velocity corresponding to a band-minimum
equals zero; yet, it can potentially have anomalous transport properties
resulting from breaking the time-reversal symmetry (i.e., violation of the
Onsager reciprocity relations)~\cite{Halperin:89}.

In summary, we have found that a long-lived non-equilibrium superfluid 
state with broken time-reversal and translational symmetries can be realized 
with cold bosons in the first excited Bloch band of a double-well optical 
lattice. The lifetime of this metastable state is shown to be several orders 
of magnitudes longer than the typical tunnelling time. Experimental realization using,
for example, stimulated Raman transitions~\cite{Mueller+:07} or
a similar method, is clearly called for.

{\it Acknowledgments}: The authors acknowledge useful discussion with
M. Anderlini and J.V. Porto, and comments by E. Zhao. This work was
supported by ARO W911NF-07-1-0293 (VMS and WVL), 
by ARO W911NF0810291 and NSF DMR-0804775 (CW), 
and by ARO-DARPA (SDS).

\bibliography{imbalance_fermi}
\bibliographystyle{apsrev}

\end{document}